\documentstyle[12pt]{article}
\newcommand{\be}{\begin{equation}}
\newcommand{\ee}{\end{equation}}
\newcommand{\ba}{\begin{eqnarray}}
\newcommand{\ea}{\end{eqnarray}}
\newcommand{\baa}{\begin{eqnarray*}}
\newcommand{\eaa}{\end{eqnarray*}}
\newcommand{\lab}[1]{\label{#1}}

\newcommand{\bhat}{\hat{\beta}}
\newcommand{\ah}{$\alpha$-helix }
\newcommand{\aten}{(Ala)$_{10}$}
\newcommand{\afif}{(Ala)$_{15}$}
\newcommand{\atwo}{(Ala)$_{20}$}
\newcommand{\vten}{(Val)$_{10}$}
\newcommand{\gten}{(Gly)$_{10}$}
\begin{document}
{\pagestyle{empty}
\rightline{ETH-IPS-95-06}
\rightline{NWU-1/95}
\rightline{March 1995}
\vskip 2.5cm
{\renewcommand{\thefootnote}{\fnsymbol{footnote}}
\centerline{\large \bf Thermodynamics of Helix-Coil Transitions}

 \vskip 0.5cm
 \centerline{\large \bf Studied by Multicanonical
             Algorithms}
}
\vskip 3.0cm

\centerline{Yuko Okamoto$^{\dagger ,}$
\footnote[1]{Address after April 1, 1995: Department of
Theoretical Studies, Institute for Molecular Science, Okazaki 444,
Japan.}
and Ulrich H.E.~Hansmann$^\#$}
\vskip 1.5cm

\centerline {$^\dagger${\it Department of Physics,
Nara Women's University, Nara 630, Japan}}
\vskip 0.5cm

\centerline{$^\#${\it  Interdisciplinary Project Center for Supercomputing
(IPS)}}
\centerline{{\it  Eidgen{\"o}ssische Technische Hochschule (ETH)
 Z{\"u}rich, 8092 Z{\"u}rich, Switzerland}}

\medbreak
\vskip 3.5cm

\centerline{\bf ABSTRACT}
Thermodynamics of helix-coil transitions in amino-acid homo-oligomers
are studied by the recently proposed multicanonical algorithms.
Homo-oligomers of length 10 are considered for three characteristic
amino acids, alanine (helix former), valine (helix indifferent), and
glycine (helix breaker).  For alanine other lengths (15 and 20)
are also considered
in order to examine the length dependence.  From one multicanonical
production run with completely random initial conformations, we have
obtained the lowest-energy conformations and various thermodynamic
quantities (average helicity, Zimm-Bragg $s$ and $\sigma$
parameters, free energy differences between
helix and coil states, etc.) as functions of temperature.  The results
confirm the fact that alanine is helix-forming, valine is
helix-indifferent, and glycine is helix-breaking.

\vfill
\newpage}
 \baselineskip=0.8cm
\noindent
{\bf INTRODUCTION} \\
Recent experimental measurements (for a review, see
Ref.~\cite{CB})
of the \ah
propensities of amino acids in short peptide systems have raised
a renewed interest in the theoretical studies of \ah formation, and
a number of simulation results have been
reported.\cite{FNKO}--\cite{YOS}

The major difficulty in conventional protein simulations such as
molecular dynamics lies in the fact that simulations at temperatures of
experimental interest tend to get trapped in
one of a huge number of local minima of potential energy surface.
Hence, the simulations strongly depend on the initial conditions.
This is why most simulations start from a folded conformation
that is suggested by X-ray or NMR experiments
and the unfolding of the conformation is studied.  However, this is
a serious limitation if one is interested in the prediction of
protein structures from the first principles without the use of
information on experimentally implied structure.  Hence, novel
algorithms that accelerate thermalization are in urgent demand.
Recently, the authors proposed the application of the multicanonical
algorithms \cite{MU,MU3} to the protein
folding problem.\cite{HO}
The performance of the algorithm was compared with that of Monte
Carlo simulated annealing,\cite{SA} another effective method for
overcoming the
above-mentioned multiple-minima problem, and it was claimed that
the former is superior to the latter.\cite{HO94_2,HO94_3}  The same
algorithm was referred
to as entropy sampling by another group,\cite{HS} but the proof of the
equivalence of the two methods was given to clarify the
matter.\cite{COMM}
Apart from the protein folding problem the multicanonical approach was
also successfully applied to the similar problem of
spinglasses.\cite{SG1}--\cite{SG3}  The advantage of this
new algorithm lies
in the fact that it not only alleviates the
multiple-minima problem but also allows the calculation of various
thermodynamic quantities as functions of temperature from
one simulation run.
The purpose of the present work is to further test the
effectiveness of the new algorithm in the study of
thermodynamics of the protein folding problem.

In this article, we study thermodynamics of helix-coil transitions
in amino-acid homo-oligomers
by multicanonical algorithms.
Preliminary results were reported elsewhere.\cite{OH95}
Homo-oligomers of length 10 are considered for three
characteristic amino acids, alanine
(helix former), valine (helix indifferent), and glycine (helix
breaker).  We first investigate the lowest-energy conformations
obtained by our simulations.  We then calculate various thermodynamic
quantities (such as the average \% helix, Zimm-Bragg $s$ and $\sigma$
parameters, free energy differences of
helix-coil transitions, etc.) over a wide range of temperatures (from
100 K to 1000 K).
To our knowledge this is the first time that such a wide range of
temperatures could be covered by a single simulation run to calculate
various thermodynamic quantities.  The Zimm-Bragg parameters are
compared with that of recent experiments.

\noindent
{\bf METHODS} \\
\noindent
{\bf Multicanonical Ensemble} \\
Although the algorithms are explained in detail elsewhere
(see, for instance Refs.~\cite{MU3,HO,HO94_3}), we briefly
summarize the idea and implementations of the method for completeness.

Most Monte Carlo (MC) simulations are done in the canonical
ensemble which
 is characterized by the Boltzmann weight factor
$
w_B (E)\ =\ \exp \left( - \bhat E \right) .
$
Here $\bhat \equiv 1/RT$ is the inverse temperature. States of energy $E$
are then distributed according to the probability
\be
P_{B}(T,E)\ ~\propto ~n(E) w_{B}(E)~,
\lab{pb}
\ee
where $n(E)$ is the density of states.  Since $n(E)$ is a rapidly
increasing function and the Boltzmann factor decreases exponentially,
$P_{B}(T,E)$ generally has a bell-like shape with its value varying
many orders of magnitude as a function of $E$.
On the other hand,
in the multicanonical approach Monte Carlo simulations are performed
in an artificial {\em multicanonical} ensemble,\cite{MU} which
 is {\it defined} by the condition that
the probability distribution of the energy shall be constant:
\be
P_{mu} (E) ~\propto ~ n (E) w_{mu} (E) = {\rm const}.
\lab{pd}
\ee
All energies have equal
weight and
a one-dimensional random walk in energy space is realized,
which insures
that the system can overcome any energy barrier.
Note that from Eq.~(\ref{pd}) we have
\be
w_{mu} (E) ~\propto ~n^{-1} (E)~. \lab{e3}
\ee
Unlike in the case for the canonical ensemble, the
multicanonical weight factor $w_{mu} (E)$ is
not {\it a priori} known, and one needs its estimator
for a numerical simulation. Hence, the multicanonical ansatz consists of
three steps:
 In the first step the estimator of the multicanonical weight factor
$w_{mu} (E)$ is calculated.
 Then one performs  with this weight factor a multicanonical
simulation with high statistics.
The standard Markov process (for instance, in a Metropolis update scheme
\cite{Metro})
is well-suited for generating configurations which
are in equilibrium with respect to the multicanonical distribution.
Monitoring the energy in this simulation one would see that
 a random walk between high energy states and ground-state
configurations is realized.
In this way information  is collected over the whole energy range.
Finally, from this simulation one can not
 only locate the energy global minimum but also obtain the
canonical distribution at any inverse temperature $\bhat$
for a wide
range of temperatures by the re-weighting techniques:\cite{FS}
\be
P_B(T,E) \propto P_{mu} (E) w^{-1}_{mu} e^{-\bhat E}~.\lab{erw}
\ee
This allows one to calculate any thermodynamic quantity
at temperature $T$.  For instance, the expectation value of a
physical quantity ${\cal O}$ at temperature $T$ is given by
\be
< {\cal O} >_T ~= \frac{\displaystyle{\int dE~ {\cal O} (E) P_B(T,E)}}
                      {\displaystyle{\int dE~ P_B(T,E)}}~.\lab{erwp}
\ee
In the ideal case a re-weighting is possible to all temperatures.
However, in reality
it may not be possible or useful to ensure the condition of
Eq.~(\ref{pd}) for all energies, but only for an interval
$E_{min} \le E \le E_{max}$~. In this case, the range of temperatures
\be
T_{min} \le T \le T_{max}
\ee
for which the re-weighting yields correct expectation values has to be
determined from the condition \cite{MU3}
\be
E_{min} \le \ <E>_{T} \ \le E_{max}~.
\label{Tlimit}
\ee

The crucial point is the first step: calculating the estimator
for the multicanonical weight factor $w_{mu} (E)$. In Ref.~\cite{HO94_3}
we proposed for this purpose the
following iterative procedure:
\begin{enumerate}
\item Perform a canonical Monte Carlo simulation at a sufficiently high
 temperature $T_0$. In our case we chose $T_0 = 1000$ K. The weight
factor
 for this simulation is given by
$ w_B (E) = e^{-\bhat_0 E} $ with $\bhat_0 = 1/R T_0$.
 Initialize the array $S (E)$ to zero, where $E$ is discretized
with bin width $\delta E$ ($=1$ kcal/mol in the present work).
\item Sample the energy distribution obtained in the previous simulation
as a
histogram $H(E)$ with the same bin width as for $S (E)$.
In the first iteration (step 1 above) determine
$E_{max}$ as the value near the mode where the histogram has
its maximum ($E_{max}$ is fixed throughout the iterations).
Let $E_{min}$ be the lowest energy obtained throughout the preceding
iterations.
For all $H(E)$ with entries greater than a certain minimum
value (say, 20)
and
$E_{min} \le E \le E_{max}$,
update the array $S(E)$ by
\be
S(E) = S(E) + \ln H(E)~.
\ee
\item Calculate the following multicanonical parameters $\alpha (E)$
and $\beta (E)$ from the array $S(E)$:
 \begin{equation}
 \beta (E) = \left\{ \begin{array}{ll}
		     \bhat_0 & ,~E \ge E_{max}\cr
                 \bhat_0 +    \frac{ \displaystyle S(E') -
                                          S(E)}{\displaystyle
		     E' - E} & ,~
                      E_{min} \le E < E' < E_{max}\cr
                     \beta (E_{min}) & ,~ E < E_{min}
                      \end{array} \right.
 \label{bcalc}
 \end{equation}
 and
 \begin{equation}
 \alpha (E) = \left\{ \begin{array}{ll}
                      0 &,~ E \ge E_{max}\cr
                      \alpha (E') + (\beta (E') - \beta (E))E' &,~ E < E_{max}
                      \end{array} \right.
 \end{equation}
 where $E$ and $E'$ are adjacent bins in the array $ S (E)$.
 \item Start a new simulation with the multicanonical weight factor
 defined by
 \begin{equation}
 w_{mu}(E) = e^{-\beta(E) E - \alpha (E)}~.\lab{ewf}
 \end{equation}
 \item Iterate the last three steps
until the obtained distribution $H(E)$ becomes reasonably
flat in the chosen energy range.
\end{enumerate}
While this method for determining the multicanonical weight factor
$w_{mu}(E)$ is quite
general, it has the disadvantage that it
requires a certain number of iterations which is not {\it a
priori} known. For the calculations in Ref.~\cite{HO} and
Ref.~\cite{HO94_3}, about 40 \% and 4 \% of the
total CPU time was respectively spent for this task.
We remark that the
above method of calculating multicanonical weights is by no means
unique. Especially it is not necessary to choose the parametrization of
Eq.~(\ref{ewf}) for the multicanonical weight factor.
However, with this parametrization and its introduction of ``effective''
temperatures $\beta(E)$
the connection to the canonical ensemble becomes apparent. Especially,
if the parameters $\beta (E)$ of Eq.~(\ref{bcalc}) are a monotone
function
of the energy $E$, then they indicate the range of temperatures,
for which a valid re-weighting is possible (see Eq.~(\ref{Tlimit})).
Note also that $S(E)$ for $\bhat_0=0$ is an estimator of
the microcanonical entropy.

\noindent
{\bf Multicanonical Annealing} \\
If one is just interested in the
ground-state structure, it may be worthwhile to use instead
a variant of the multicanonical
method, {\it multicanonical annealing}. \cite{LC,HO94_2,HO94_3}
 Here, an upper bound in energy is introduced at
 the other end of the annealing direction, rejecting all
 attempts beyond this bound. Annealing is achieved by moving the bound in
 the annealing direction while keeping the sampling interval $\Delta E$
 fixed.  Within this energy  interval the system
 can move out of local minima as long as their barrier heights do not
  exceed the upper limit of the energy range.
 Because of the finite interval size $\Delta E$, the
 MC procedure will no longer be
 ergodic and it is not possible to find the equilibrium properties of the
 system. Hence, unlike in the case for regular multicanonical algorithm,
 the canonical distribution cannot be reconstructed. For the
 purpose of annealing this does not matter as long as one chooses the
 sampling interval large enough to allow important fluctuations throughout
 the annealing process. However, since ergodicity is not fulfilled, one has
 to repeat the annealing process several times with different initial
 conformations
 to make sure that one has found a good approximation to the
 global minimum.
 The optimal sampling interval
 $\Delta E$ is
 not known {\it a priori} for multicanonical annealing and has
 to be chosen
 on a trial-and-error basis.

The multicanonical annealing algorithm is discussed in detail in Refs.~
\cite{HO94_2,HO94_3} where first tests of the method for the
protein folding problem were performed.
 Results  better than those obtained
 by simulated annealing were reported. The following
implementation of the algorithm was proposed:
\begin{enumerate}
\item Perform a short canonical Monte Carlo simulation at a
sufficiently high
 temperature $T_0$. Again we chose $T_0=1000$ K in the present work.
 Initialize an array $S (E)$ to zero, where $E$ is discretized
with bin width $\delta E$ ($=1$ kcal/mol in the present work).
\item Sample the energy distribution obtained in the previous
simulation as a histogram
 $H(E)$.
Let $E_{min}$ be the lowest energy obtained throughout the preceding
iterations.
For all $H(E)$ with entries greater than a certain minimum
value (say, 20),
update the array $S(E)$ by
\be
  S(E) = S(E) + \ln H(E) ~.
\ee
\item Update the upper bound $E_{wall}$ of the sampling interval by
\begin{equation}
  E_{wall} = \max (E_{last},E_{min} + \Delta E)~,
\end{equation}
where $\Delta E $ is the  size of the sampling energy range and $E_{last}$
the energy of the last conformation.
\item Calculate the following parameter $\beta_{min}$ by
\begin{equation}
\beta_{min} =\frac{ \displaystyle S (E_{wall}) - S (E_{min})}
{E_{wall} - E_{min}}~.
\end{equation}
\item Define the new weight factor by
\begin{equation}
 w(E) = \left \{ \begin{array}{ll}
                 0 &,~ E > E_{wall}\cr
                 e^{-S(E)} &,~ E_{min} \le E \le E_{wall}\cr
		     e^{-S(E_{min}) - \beta_{min} (E - E_{min})} &,~ E < E_{min}
                 \end{array} \right.
\end{equation}
and perform a new simulation with this weight factor, starting from
the last conformation of the preceding simulation.
\item Iterate the last four steps till no newer $E_{min}$ is
found for a certain number of consecutive iterations.
\end{enumerate}

\noindent
{\bf Peptide Preparation} \\
We considered amino-acid homo-oligomers of Ala, Val, and Gly. By experiments
Ala is known to be a strong helix former, while Val and Gly are known to
be helix indifferent and helix breaker, respectively.
The number of residues, $N$, for each homo-oligomer was taken to be 10.
For (Ala)$_{N}$, however, the cases for $N=15$ and 20 were also
considered in order to examine the $N$ dependence.
Since the charges at peptide termini are known to
reduce helix content,\cite{IOT,SKY}
we removed them by taking a neutral NH$_2$-- group at the
N-terminus and a neutral --COOH group at the C-terminus.

\noindent
{\bf Potential Energy Function} \\
The potential energy function
$E_{tot}$ that we used is given by the sum of
the electrostatic term $E_{C}$, 12-6 Lennard-Jones term $E_{LJ}$, and
hydrogen-bond term $E_{HB}$ for all pairs of atoms in the peptide together with
the torsion term $E_{tor}$ for all torsion angles:
\begin{eqnarray}
E_{tot} & = & E_{C} + E_{LJ} + E_{HB} + E_{tor},\\
E_{C}  & = & \sum_{(i,j)} \frac{332q_i q_j}{\epsilon r_{ij}},\\
E_{LJ} & = & \sum_{(i,j)} \left( \frac{A_{ij}}{r^{12}_{ij}}
                                - \frac{B_{ij}}{r^6_{ij}} \right),\\
E_{HB}  & = & \sum_{(i,j)} \left( \frac{C_{ij}}{r^{12}_{ij}}
                                - \frac{D_{ij}}{r^{10}_{ij}} \right),\\
E_{tor}& = & \sum_l U_l \left( 1 \pm \cos (n_l \chi_l ) \right).
\end{eqnarray}
Here, $r_{ij}$ is the distance between the atoms $i$ and $j$, and
$\chi_l$ is
the torsion angle for the chemical bond $l$.
The parameters ($q_i,A_{ij},B_{ij},C_{ij},
D_{ij},U_l$, and $n_l$) for the energy function were adopted
from ECEPP/2.\cite{EC1}--\cite{EC3}  Since one can avoid the
complications of electrostatic and hydrogen-bond interactions of
side chains with the solvent for nonpolar amino acids, explicit
solvent molecules were neglected
and the dielectric constant $\epsilon$ was set equal to 2.
This is a crude approximation,
and we have
to keep this limitation in mind when we compare our simulation results
with experiments.
The computer code
KONF90 \cite{KONF,KONF2} was modified to accommodate the multicanonical
algorithms.
(There are slight differences in conventions between KONF90 and
the original version of ECEPP/2;
for example, $\phi_1$ of KONF90 is equal to
$\phi_1 - 180^{\circ}$ of ECEPP/2, and energies are also different by
small irrelevant constant terms.)
The peptide-bond
dihedral angles $\omega$ were fixed at the value 180$^\circ$
for simplicity,
which leaves $\phi_i,~\psi_i$, and $\chi_i$ ($i=1, \cdots, N$)
as independent degrees
of freedom.  Since Ala has one $\chi$ in the side chain,
Val three $\chi$'s, and Gly none, the numbers of independent degrees
of freedom
are 30, 50, 20, 45, 60 for \aten, \vten, \gten, \afif, \atwo,
respectively.

\noindent
{\bf Computational Details} \\
One MC sweep updates every
dihedral angle (in both the backbone and the side chains) of the
homo-oligomers once.

For regular multicanonical simulations, the multicanonical
weight factors were determined by the iterative procedure described
above. We needed between 40,000 sweeps (for \aten) and 100,000 sweeps
(for \vten)
for their calculation.
 All thermodynamic quantities were then  calculated from
 one production run of 200,000 MC sweeps following additional 10,000
sweeps for equilibration.

For multicanonical annealing, we performed 10 simulation runs from
different random initial conformations
with 20,000 MC sweeps (so that the total number of MC sweeps is equal to
200,000 for each homo-oligomer).  We divided the
20,000 MC sweeps of each run
in 10 annealing iterations of 2,000 MC
sweeps.
The value of the sampling interval $\Delta E$ was chosen to be
$\Delta E=15$ kcal/mol.

In all cases, each simulation started from a completely random initial
conformation (``Hot Start'').
 We remark that for the regular multicanonical algorithm,
simulations with initial conformations of an ideal helix (``Cold Start'')
were also
performed, and we found that the results are in agreement with
those from random initial conformations.
This suggests that thermal
equilibrium has been attained in our simulations.

\noindent
{\bf RESULTS AND DISCUSSION} \\
{\bf Lowest-Energy Conformations} \\
We first investigate the lowest-energy conformations obtained
from our simulations.
Here, we are able to cross-check our results of regular multicanonical
simulation by those of multicanonical annealing.
The criterion we adopt for $\alpha$-helix formation is as
follows:\cite{YOP}
We consider
that a residue is in the right-handed $\alpha$-helix configuration
when the
dihedral angles ($\phi,\psi$) fall in
the range ($-70 \pm 20^{\circ},-37 \pm 20^{\circ}$).
The length $\ell_R$ of a helical segment is then defined by the
number of successive
residues which are in the right-handed $\alpha$-helix configuration.
The number $n_R$ of
helical residues in a conformation is defined by the sum of $\ell_R$
over all right-handed helical segments in the conformation.
Note that we have $n_R \ge \ell_R$ with the equality holding when
there exists
just one (or no) helical segment in the conformation.
Since Gly has no
side chain, it can produce both right-handed and left-handed helices.
Analogous quantities $\ell_L$ and $n_L$ for left-handed \ah are
defined with obvious reversal of signs for the dihedral angles.

The conformations obtained during the simulation
are classified into two states, helix and coil.
Here, a conformation is considered to be in the helix state,
if it has a segment with helix length $\ell \ge 3$.
A conformation is considered to be in the coil state, if it is not
in the helix state.
We remark that this definition of the helical conformation is in
some sense arbitrary; for instance, we could define the helical
conformation as that with $\ell \ge 2$ instead.  But
one has to draw a line somewhere, and we chose this definition, since
$\ell = 3$
corresponds to roughly one turn of $\alpha$-helix.
In a similar way, the above definition
for the helical residue is also by no means unique.
We have chosen the central values of $-70^{\circ}$ and $-37^{\circ}$
here,
because they were  the
average values of $\phi$ and $\psi$ in the helical residues we
obtained with  KONF90 in previous simulations.\cite{YOP}
We have to keep in mind these ambiguities in the definitions of
helical residue and conformation, but we remark that we found no
qualitative changes
when we checked our results
by using different definitions in our analysis.

In Table~1 we list the energy $E$ (kcal/mol) and the
maximum helix lengths
$\ell_R$ and $\ell_L$ of the lowest-energy conformations obtained during
each of 10 multicanonical annealing runs of 20,000 MC sweeps for
(Ala)$_{10}$, (Val)$_{10}$, and (Gly)$_{10}$.  Each run was
performed from a completely random initial conformation. Hence, the results
 of the 10 runs are independent of each other.
These quantities are listed separately for
conformations in both helix and coil states.
Our aim was to study if there is a unique ground state
 or a multitude of low-energy states. Despite the experimentally
observed
differences in $\alpha$-helix propensities, one could assume that an
ideal helix is the unique ground-state structure for all
homo-oligomers, since this
structure is energetically favored by the van der Waals term.
The observed
propensity differences would then be caused by the entropic effects.
To check this conjecture we also calculated
the minimized energies of an idealized helical conformations for the
three homo-oligomers
and compared them with our numerical results. To obtain the idealized
helical structure we first set all backbone angles to
$\phi = -70^{\circ}$ and
$\psi =-37^{\circ}$ (i.e., $\ell = 10$) and the side chain
angles $\chi$ to $60^{\circ}$ for
\aten \ and $180^{\circ}$ for \vten. The energies of these initial
conformations were then minimized by the Newton method.
The resulting energies are also listed in Table~1.
 In Table~2 we give the dihedral angles of the
 lowest-energy conformations of Table~1 for completeness.

The first thing one can tell from Table~1 is that
the lowest-energy
structures of \aten \ are all right-handed $\alpha$-helix with almost
100 \% helicity ($\ell_R = 8$).
The energy of these conformations is
comparable to that of the minimized ideal helical conformation.
Note that the energy
of the minimized ideal-helix structure is slightly higher than the
lowest energy of our multicanonical annealing simulation.
 We conclude that the structure
with the global-minimum energy for \aten \ is an ideal helix and that there
is a continuum of excited states whose energy and structure are
similar to
those of the ground state.
 We remark that with the present
multicanonical annealing simulation, the probability of finding
the ground-state structure ($\ell_R \ge 8$) is 60 \%; 6 out of 10
runs found it.  The probability would be higher if we increased the
statistics of each run.
 Finally, we found that
the side chain
structure of \aten \ is also unique for the ground-state structure;
namely, the values of $\chi$ are one of $60^{\circ}$, $-60^{\circ}$,
and $180^{\circ}$, which are all equivalent angles because of the
three-fold rotational symmetry of the alanine side chain (see Table~2).

Things look different for the other two homo-oligomers.
 Since Gly has no
 side chain, it can produce both right-handed and left-handed helices.
The longest left-handed helical length
$\ell_L$ for \gten \ is as large as 6 in Table~1.
On the other hand, the longest right-handed helical length
$\ell_R$ for \gten \ is only 3 in Table~1.  Here, it appears that
with our energy function, left-handed helix is favored over right-handed
one.  However, as is discussed below, the regular multicanonical
production run did produce longer right-handed  helix ($\ell_R =5$)
with similar energy (see Table~3 below).  Hence, this tendency
is weak and caused by the lack of statistics in our multicanonical
annealing simulations.  The energy of the lowest-energy helical
conformation is higher than or
comparable with that of coil structures.  The lowest-energy
structure is a coil (run 8).  Furthermore, the
energy of the minimized ideal-helix structure is also higher than that
of the lowest-energy coil structure. Similar observations were
made in Ref.~\cite{Hagai}
and it was conjectured that in a short $\alpha$-helix the
electrostatic interaction is unfavorable
due to parallel arrangement of peptide bond
dipoles.\cite{Hagai}  However, it was argued there that for
larger chains, the
attractive van der Waals term will win and a helix conformation would
be the ground state. From our multicanonical annealing results we can
only conclude that for \gten \  $\alpha$-helix is {\it not} the
ground-state conformation and that there exist a multitude of
different low-energy (coil) states.
Without side chains, \gten \ is so flexible that the ground state
cannot be strongly
energetically favored compared to other structures.
In this sense  \gten \ is like a
 spinglass with many local minima of energy near the global minimum.

For
\vten, some of the low-energy structures are helical and others not.
Our lowest-energy conformation (run 9) is the one with the highest
value of $\ell_R ~(=6)$.
As is shown in Table~1, the lowest-energy conformations obtained
by the multicanonical annealing runs have a much higher
energy (at least 6 kcal/mol) than the minimized ideal-helix structure.
Hence, we have to admit that the multicanonical annealing simulation
did not find the energy global minimum for \vten, although it got
close to it.
 We conjecture that for \vten\  an ideal helix structure is
the ground state and that
because of the steric hindrance of the side chains (Val has a large
side chain),
it is separated by high energy barriers from a multitude of states
with smaller helical lengths and slightly higher energy.  Hence,
it is very difficult to reach the ideal helix structure from a completely
random initial conformation.

We now compare the above results by multicanonical annealing with
those by regular multicanonical algorithm production runs.  For
each homo-oligomer, \aten, (Ala)$_{15}$, (Ala)$_{20}$, \vten,
and \gten, one production run with
200,000 MC sweeps (after 10,000 sweeps for equilibration)
was made.  Hence, the statistics is 10 times more
than one multicanonical annealing run
 (but 10 runs were made for the
 latter).
In Table 3 we list the energy
$E$ (kcal/mol),
helix lengths $\ell_R$ and $\ell_L$, and dihedral angles of the
lowest-energy
conformations obtained by each multicanonical production run.
These quantities are again listed separately for the lowest-energy
conformations in both helix and coil states.  As is clear from the
table, the uniqueness of the global-minimum state (ideal helix
structure) for \aten \ is apparent in accord with
the implication of the multicanonical annealing runs; we have roughly
the same energy ($\approx -10$ kcal/mol), helix length
($\ell_R = 8$), and dihedral angles in Tables 2 and 3.
We find that even the structure of the termini is unique;
the dihedral angles $\psi_1$, $\phi_{10}$, and $\psi_{10}$ are
essentially the same in Tables~2 and 3.
We remark that the lowest-energy structures for (Ala)$_{15}$ and
(Ala)$_{20}$ obtained by the multicanonical production runs were
also ideal helix with helix lengths $\ell_R = 13$ and 18, respectively.
Furthermore, the values of $\chi$ were one of $60^{\circ}$,
$-60^{\circ}$, and $180^{\circ}$.
Non-uniqueness of the lowest-energy structures for \vten \ and \gten \
are also clear in the table; we have different helix length and
the dihedral angles but similar energies ($\approx 2, -8$ kcal/mol
for \vten, \gten, respectively) in Tables 2 and 3.
Again, the global-minimum conformation for \vten\ (ideal helix)
was not reached.  However, we remark that the lowest energies for helix states
reached by multicanonical production runs for \aten\ and \gten\ are
less than the minimized ones in Table~1.  In Figure 1 we show
stereoscopic views of the lowest-energy structures from Tables 2
and 3 for completeness.  The lowest-energy left-handed $\alpha$-helix
structure for (Gly)$_{10}$ (from Table 3) is also shown in the figure.

As one can read off from Table 3, the differences
$\Delta E \equiv E_C - E_H$ of the lowest energies between coil state
($C$) and helix state ($H$) are 12.4, 4.8, $-0.9$ kcal/mol for \aten,
\vten, \gten, respectively.  (Note that these values are lower bounds
and the more realistic values may be even larger, since
their coil states in Table~3 are almost helical.  Namely, if we define
a helical residue by the condition
$(\phi,\psi)=(-70 \pm 23^{\circ},-37 \pm 23^{\circ})$ instead,
these structures are considered to be helical.)
This and our results from the multicanonical
annealing runs imply that the helix state is
energetically favored strongly for \aten \ and slightly favored for
\vten, while it is not favored for \gten.\cite{YOP}  These observed energy
differences
are a natural explanation for the experimentally observed fact
 that Ala is a helix former and Gly is a helix
breaker, while Val comes in between the two.  More quantitative
arguments about this point will be given in the following subsections.

\noindent
{\bf Energy Distribution and Specific Heat}\\
While multicanonical
annealing gives information only on the lowest-energy state,
regular multicanonical algorithm also allows the calculation of
thermodynamic quantities at various temperatures. The range of
temperatures, for which a valid re-weighting is possible, is given
in Eq.~(\ref{Tlimit}). Since we are fairly sure that we determined the
ground state for \aten, we can in principle trust our results
here down to $T= 0 $~K. However, we do not know for sure that
we found the ground state
for \gten\  and we know that we missed it for \vten. Hence, we had to
determine in these cases the range of temperatures that allow
reliable re-weighting by Eq.~(\ref{Tlimit}).
As an upper
limit to which we can trust our results for the three homo-oligomers,
we found $T_{min} = 200$ K. Therefore we restricted our analysis to
temperatures  between 200 K and 1000 K in most of the cases below.

In Figure 2 we show 3-dimensional plots of the probability distribution
of energy as a function of temperature for \aten, \vten, and \gten.
These results were obtained from the same multicanonical
production runs as above
with the reweighting techniques of Eq.~(4).  At
each fixed temperature $T$, the probability distribution
$P_B (T,E)$
corresponds to a canonical distribution at this temperature.  Thus,
it is given by Eq.~(1) and has a bell-shape.  The width of the
bell-shape is large for high temperatures reflecting the large
energy fluctuations.  As the temperature decreases, this width
decreases but the height of the
peak increases (since the probability is normalized to 1),
and it should behave as a $\delta$-function
$\delta (E - E_0)$ in the limit $T \rightarrow 0$ K, where $E_0$ is the
global-minimum energy.  These properties of canonical ensemble
are indeed clearly seen in Figure 2.

In order to demonstrate the reliability and superiority of our method,
we also performed
for (Ala)$_{10}$ and (Ala)$_{20}$
canonical simulations of 200,000 sweeps at temperature $T = 270$ K and
compared the obtained distribution of energies with the one from
multicanonical simulation by reweighting. Again 10,000 sweeps were
performed for equilibration of our system. Figures 3 show our results.
Even for $N=10$ we observe remarkable differences in the shape of the
distribution, and they are totally different for $N=20$. In
the case of $N=10$ we observe for the canonical simulation
a pronounced tail for energies larger than
1 kcal/mol while both distribution look similar for lower energies. When
we looked deeper into the time series for the canonical run, we found that our
system need at least 80,000 sweeps more than the estimated 10,000 sweeps for
equilibration. When we discard these sweeps, the distribution
agreed with that obtained by the multicanonical production run.
In the case for $N=20$ the system did not equilibrate
at all in the canonical simulation. On the other hand, the
multicanonical simulation was fully
equilibrated: The
distribution did not depend on the initial conformations. Moreover, our
method gives information on a whole temperature range, not only
a single one. This
demonstrates the superiority of our method.

In Figure 4 we show the \lq \lq specific heat" as a function
of temperature
for the three homo-oligomers, \aten, \vten, and \gten.  The specific
heat here is defined by the following equation:
\begin{equation}
  C(\bhat)  = {\bhat}^2 \ \frac{<E_{tot}^2> - <E_{tot}>^2}{N},
\end{equation}
where $\bhat$ is the inverse temperature $1/RT$ and $N \ (=10)$ is the
number of residues in the oligomers.  It was  calculated
from the reweighting techniques of Eqs.~(4) and (5).  The data in the figure
all
represent a peak at a certain temperature, indicating that there is
a crossover between a coil phase and a helix phase.
The temperatures
at the peak, transition temperatures, are
$T_c \approx 430$, 330, 360 K for \aten, \vten, \gten,
respectively.
The peak structure is most conspicuous for \aten, and this
corresponds to a
transition from a random-coil phase to an ideal-helix phase.
The transition temperature $T_c$ for \aten \ is
rather high.  This indicates that \aten \
is substantially helical (see Figure 7 below) even at temperatures
near 400 K.
The peaks in specific heat are not as sharp for
\vten \ and \gten, and this is
a typical characteristic of a spinglass.

To study if the mentioned crossover is an indication for a
``phase transition'' we show in Figure 5
 the specific heat as a function of temperature
for \aten, \afif, and \atwo.  As the number $N$ of residues increases,
the peak temperature and peak height also increase, and the peaks become
more pronounced, suggesting the divergence of the specific heat in
the thermodynamic limit ($N \rightarrow \infty$).  This behavior would
be expected for a phase transition.
We do not expect similar behavior for the other
homo-oligomers, but did not study this question because of limited
computer time.

Finally, we investigate how each of the energy terms in
Eqs.~(16) -- (20)
varies as $T$ changes.
In Figure 6 we show each energy term as a function of
temperature for \aten, \vten, \gten, and \atwo.  In all
cases, each term monotonically increases as $T$ increases.
The changes, however, are very small except for the Lennard-Jones
term, $E_{LJ}$, indicating that $E_{LJ}$ is the key factor for the
folding of these homo-oligomers.\cite{YOP}
For (Ala)$_N$ ($N=10$ and 20)
we observe a steep change of $E_{tot}$ and $E_{LJ}$ (and other
terms) at the transition temperatures ($T= 400 \sim 500$ K),
and the slope
becomes steeper as $N$ increases.  These facts are reflected in
the pronounced peaks in the specific heat in Figure 5 and again
suggest the existence of a phase transition for this polymer.
Work is under progress to study the nature of this transition in
more detail.

\noindent
{\bf Thermodynamics of Helix-Coil Transitions} \\
In Figure 7 we show the average helicity ${< n >} \over N$ as a function
of temperature for \aten, \vten, and \gten.  These were again calculated
by the reweighting techniques of Eqs.~(4) and (5).
Here, $<n>$ stands for $<n_R>$.  From the regular multicanonical
production run, it was found that
right-handed and left-handed helicities are essentially equal
for \gten :
$<n_R> \approx <n_L>$.  The average helicity
tends to decrease monotonically as the temperature increases because
of the increased thermal fluctuations.  Around the room temperature,
\aten \ is substantially helical ($\approx 75$ \% helicity),
\vten \ is slightly helical ($\approx 30$ \% helicity), and \gten \
is hardly helical ($< 10$ \% helicity).  This is consistent with the
fact that Ala is a helix former and Gly is a helix
breaker, while Val comes in between the two.
For reasons described above we could not reweight our $<n>$ data to
$T = 0$ K
and we did not dare to extrapolate them from the safe temperature range
to $T = 0$ K. However, the qualitative behavior of $<n>$ for decreasing
temperature supports our conjecture that the ground state of both
\aten\  and \vten\ is an ideal helix, while this is not the case
for \gten.

Again we investigated for (Ala)$_N$ the dependence of this quantity
on the number of residues $N$.
In Figure 8 we show ${<n>} \over N$ as a function of temperature for
\aten, \afif, and \atwo.  The data indicate that the longer the
homo-oligomer, the more helix-forming it is, as we guessed above.
  The difference between
the residue numbers 15 and 20 is smaller than that with 10, indicating
that the $N \rightarrow \infty$ limit for (Ala)$_N$ may be reached already near
$N=20$.

According to the Zimm-Bragg model,\cite{ZB}  the average number
of helical residues $<n>$ and the average length $<\ell>$ of a helical
segment are given for large $N$ by
\begin{eqnarray}
{{<n>} \over N}~ &=& ~{1 \over 2} - {{1-s} \over {2
\sqrt{(1-s)^2 + 4s \sigma}}}~, \\
<\ell>~~ &=& ~1 + {2s \over {1-s+\sqrt{(1-s)^2 +4s \sigma}}}~,
\end{eqnarray}
where $N$ is the number of residues, and $s$ and $\sigma$ are the
helix propagation parameter and nucleation parameter, respectively.
Note that $s \ge 1$ implies that ${{<n>} \over N} \ge {1 \over 2}$
(more than 50 \% helicity).
 From these equations with the values of ${<n>} \over N$ and $<\ell>$
calculated from the multicanonical production runs,
one can obtain estimates of $s$ and $\sigma$
parameters.  The values at six temperature values $T=250,$ 300, 350,
400, 450, and 500 K for \aten, (Ala)$_{15}$, (Ala)$_{20}$,
\vten, and \gten \ are listed in Table~4.
The $s$ parameters monotonically decrease
as the temperature increases because the increased thermal fluctuations
will decrease helical content as the temperature increases (see
Figure~7).  Note that the difference between $N=15$ and 20 for
(Ala)$_N$ is again small, suggesting that $N=20$ is already large
enough.  The difference between $N=10$ and 20 is also within the errors.
The $s$ parameters at experimentally relevant temperature
($\approx 270$ K) are about 1.6, 0.6, 0.2 for Ala, Val, Gly,
respectively.  These are in good agreement with the experimental
results,\cite{CB} where they give $s$(Ala) = $1.5 \sim 2.19$,
$s$(Val)= $0.20 \sim 0.93$ and $s$(Gly) = $0.02 \sim 0.57$.
 The $\sigma$ parameters in Table~4, on the other hand, tend
to be order of magnitude larger than the commonly assumed values
($\approx 10^{-3}$).  Since the data for Ala have smaller errors and
are more reliable than those for Val and Gly, we show the
$s$ and $\sigma$ values for Ala as functions of temperature in
Figure 9.
As discussed above, \aten \ has a
clear phase-transition signal with a pronounced peak in
specific heat at the transition temperature $T_c$ of $\approx 430$ K
(see Figure~4).  Above the transition temperature $T_c$,
\aten \ is in the random-coil phase, and below that
temperature it is in the helix phase (see also Figure~7).  This
transition temperature $T_c$ can also be identified with as the
temperature where $s=1$ holds (i.e., 50 \% helicity) in Figure 9a,
which is about 410 K. The small disagreement is due to the arbitrariness
in our definition of a helical state.
 As is clear from Figure 9b,
in the helix phase ($T<T_c$) the $\sigma$ parameter for Ala is
small and constant, but in the random-coil phase ($T>T_c$)
$\sigma$ starts to grow as temperature increases.  This growth of
$\sigma$ values reflects the increased thermal fluctuations that
prevent the formation of a long helix.  That is, below $T_c$
cooperativity for helix-formation wins over thermal fluctuations
but above $T_c$ thermal fluctuations win and no long helices can be
formed.

Figure 9 implies that as the number of residues $N$
increases for (Ala)$_N$, $s$ values at a fixed temperature
tend to increase and $\sigma$ values tend to decrease.
Eqs. (22) and (23) are exact in the limit
$N \rightarrow \infty$.  Hence, the $s$ and $\sigma$ should be
extrapolated in this limit. However, we found that the differences in the
 value of the
$s$ parameter at experimentally relevant temperature,
around 273 K, are within the errors. Therefore
we conjecture that $N=10$ is already large enough for
determining this quantity. We expect that systematic errors due to our
poor approximation of solvent effects are larger than the finite-size
effects. The situation may be different for  $\sigma$. This quantity is
 decreasing as
$N$ increases and is
$<0.06$ for $N=20$, which is a more acceptable value. Obviously the
reliability of our values for this quantity is much more limited by
the small number of residues in our simulations.

Another way to investigate the limitations of our results due to finite
number of residues is to look into the end effects of the homo-oligomers.
Our analyses so far have been neglecting them.
As for helical content, one expects to see fraying
at the edges.  In Figure 10 we show the percent helicity $<n> \over N$
 as a function of the residue number at
$T=250$ and 350 K for \aten, \vten, and \gten.  Again these
results were calculated by the reweighting techniques
of Eqs.~(4) and (5).
We do observe fraying for all cases.  The contrast is
most outstanding for \aten \ because it has high helicity.  The increase
of fraying as the temperature is raised is clearly seen for
\aten.

The helix-coil transition can be further studied by calculating
the free energy differences $\Delta G \equiv G_H - G_C$,
enthalpy differences $\Delta H$, and
entropy differences $T \Delta S$ between helix ($H$) and coil
($C$) states.
Here, a conformation is considered to be in the helix
state if it has a segment with helix length $\ell \ge 3$.
The free energy differences were calculated from
\begin{equation}
\Delta G = - RT \ln { {<N_{H}>} \over {<N_{C}>}} ,
\end{equation}
where $<N_{H}>$ and $<N_{C}>$ are the average numbers of
conformations in
helix state and in coil state, respectively.
The enthalpy differences were estimated from
\begin{equation}
\Delta H = <E_H> - <E_C> ,
\end{equation}
where $<E_H>$ and $<E_C>$ are average total potential energies
$<E_{tot}>$
(see Eq.~(16)) in helix and
coil states, respectively.
Finally, the entropy differences were obtained from
$\Delta G$ and $\Delta H$ by the relation
\begin{equation}
T\Delta S = \Delta H - \Delta G~.
\end{equation}

In Table 5 we list $\Delta G$, $\Delta H$, and $T \Delta S$
at six temperatures $T=250,$ 300, 350,
400, 450, and 500 K for
\aten, (Ala)$_{15}$, (Ala)$_{20}$, \vten, and \gten.
$\Delta G$ is
negative for \aten \ if $T < 450$ K.
Hence, \aten \
favors a helix state for $T < 450$ K.
Again this limiting
temperature is around the transition temperature $T_c$ of
the phase transition from the helix phase to the random-coil phase
discussed above.
For \vten, a helix state is slightly favored if $T < 300$ K, and
for \gten, a coil
state is favored at all temperatures. The reason for the variance in
helix-propensities becomes clear by looking at the enthalpy differences
$\Delta H$ for the different oligomers. For \aten\ we observe large
enthalpy differences between helix and coil states which favor the
helix state and win over the entropic term of opposite sign.
 On the other hand, the helix state is only slightly
energetically favored for \vten\  and not for \gten. For \vten\ enthalpy
and entropy differences are of the same order,
while for \gten\ the entropic term wins
and coil states are favored.
We remark that the absolute values of $T \Delta S$ become larger
as $T$ increases because of the increased thermal fluctuations.

\noindent
{\bf CONCLUSIONS} \\
In the present work we have demonstrated the effectiveness of
multicanonical algorithms and presented various quantities one
can calculate by this method.  This was done by taking the example
of helix-coil transitions of amino-acid homo-oligomers.  Our results
show that we are able to calculate
the helix propensities of amino acids and compare them with experiments.
However, our results have to be taken as preliminary. They are both
hampered by the small number of residues and even more the neglecting
of solvent effects. We expect to obtain an even better agreement
with experiments,
once we incorporate more realistic energy functions with solvent
effects included.

\vspace{0.5cm}
\noindent
{\bf Acknowledgements}: \\
Our simulations were
performed on   a cluster of fast RISC workstations at SCRI (The Florida
State University, Tallahassee, USA) and
NEC SX-3/34R at Institute for Molecular Science, Okazaki,
Japan.
This work is
supported, in part, by the Department of Energy, contract DE-FC05-85ER2500,
by a Grant-in-Aid for Scientific Research from the
Japanese Ministry for Education, Science and Culture, by the
Schweizerische Nationalfonds (Grant 20-40'838.94), and by MK
Industries, Inc. \\


\newpage
\noindent
{\Large Table Captions:}\\
{Table~1: Energies $E$ (in kcal/mol) of the lowest-energy conformations
in helix state and coil state obtained during each run of multicanonical
annealing simulation.  R-Helix and L-Helix stand for
right-handed $\alpha$-helix state and left-handed $\alpha$-helix
state, respectively.  Helix length $\ell$ is also given
for the conformations in helix state.  The null entry implies that
there was no conformation in that state with $E<35$ kcal/mol. MIHC stands
for minimized idealized helical conformation.}\\
{Table~2: Dihedral angles (in degrees) of the lowest-energy conformations
in helix state and coil state obtained in 10 runs of multicanonical
annealing simulation.  R-Helix and L-Helix stand for
right-handed $\alpha$-helix state and left-handed $\alpha$-helix
state, respectively.  The symbols * and \# indicate that the
corresponding residue is respectively in right-handed and left-handed
helix configuration.}\\
{Table~3: Dihedral angles (in degrees) of the lowest-energy conformations
in helix state and coil state obtained in multicanonical production runs.
R-Helix and L-Helix stand for
right-handed $\alpha$-helix state and left-handed $\alpha$-helix
state, respectively.  The symbols * and \# indicate that the
corresponding residue is respectively in right-handed and left-handed
helix configuration.}\\
{Table~4: Average number of helical residues $<n>$,
average length of a helical segment
$<\ell>$, and the Zimm-Bragg
$s$ and $\sigma$ parameters as functions of temperature $T$ (K).
The numbers in
parentheses represent errors.}\\
{Table~5: Free energy differences $\Delta G (\equiv G_H - G_C)$,
enthalpy differences $\Delta H$, and entropy
differences $T\Delta S$
(all in kcal/mol) between helix ($H$) and coil ($C$) states
as functions of temperature $T$ (K). The numbers in
parentheses represent errors.}\\

\newpage
Table~1.\\
\begin{table}[h]
\begin{center}
{\small
\begin{tabular}{c|cc|c|cc|c|cc|cc|c} \hline
& \multicolumn{3}{c|}{(Ala)$_{10}$}&
  \multicolumn{3}{c|}{(Val)$_{10}$}&
  \multicolumn{5}{c}{(Gly)$_{10}$}\\ \hline
State& \multicolumn{2}{c|}{R-Helix}&Coil&
       \multicolumn{2}{c|}{R-Helix}&Coil&
       \multicolumn{2}{c|}{R-Helix}&
       \multicolumn{2}{c|}{L-Helix}&Coil\\ \hline
   & $E$ & $\ell_R$ & $E$ &
     $E$ & $\ell_R$ & $E$ &
     $E$ & $\ell_R$ &
     $E$ & $\ell_L$ & $E$
\\ \hline
Run&     &          &     &
         &          &     &
         &          &
         &          &
\\
1  & $-9.1$ & 8 & 3.7 & 31.6 & 3 & 19.8 & 4.3 & 3 & & & $-5.3$ \\
2  & 15.9 & 4 & 2.5 & 6.6 & 3 & 5.9 & 7.4 & 3 & 5.1 & 3 & $-4.8$ \\
3  & $-8.9$ & 8 & 6.0 & 12.9 & 3 & 14.8 & 6.2 & 3 & 7.1 & 5 & $-5.3$ \\
4  & $-9.1$ & 8 & 11.8 & 5.3 & 3 & 6.1 & 3.2 & 3 & 16.2 & 3 & $-5.5$ \\
5  & $-8.3$ & 8 & 9.8 &  &   & 5.9 & 8.2 & 3 & 24.2 & 3 & $-4.6$ \\
6  & & & $-1.5$ & 4.6 & 4 & 6.4 & 0.1 & 3 & 12.5 & 3 & $-7.4$ \\
7  & 0.2 & 3 & $-0.9$ & 14.4 & 3 & 8.7 & 7.0 & 3 & 2.3 & 3 & $-3.6$ \\
8  & $-1.8$ & 3 & 1.5 & 4.3 & 3 & 5.1 & 5.5 & 3 & $-1.3$ & 5 & $-8.0$ \\
9  & $-9.1$ & 8 & 3.8 & 2.6 & 6 & 7.3 & 9.9 & 3 & 5.5 & 3 & $-6.7$ \\
10  & $-8.2$ & 8 & 7.0 & 7.6 & 3 & 4.0 & 27.3 & 3 & $-6.0$ & 6 & $-5.4$ \\
\hline
MIHC & $-8.8$ & 8 & & $-3.4$ & 8 & & $-3.9$ & 8 &-3.8 & 8 & \\
\hline
\end{tabular}
}
\end{center}
\label{tab1}
\end{table}

\newpage
Table~2.\\
\begin{table}[h]
\begin{center}
{\small
\begin{tabular}{cccccc} \hline
\multicolumn{6}{c}{(Ala)$_{10}$}\\ \hline
Run&\multicolumn{2}{c}{State}&
\multicolumn{3}{c}{$E$ (kcal/mol)} \\
9&\multicolumn{2}{c}{R-Helix}&
\multicolumn{3}{c}{$-9.1$} \\
\hline
Residue & $\phi$ & $\psi$ & $\chi$ & & \\
1 & 102 & 150 & $-61$ & & \\
2* & $-68$ & $-32$ & $-58$ & & \\
3* & $-69$ & $-39$ & $-62$ & & \\
4* & $-71$ & $-35$ & $-175$ & & \\
5* & $-71$ & $-36$ & 180 & & \\
6* & $-73$ & $-34$ & $-179$ & & \\
7* & $-71$ & $-36$ & $-54$ & & \\
8* & $-72$ & $-37$ & $-62$ & & \\
9* & $-71$ & $-40$ & 61 & & \\
10 & $-145$ & 100 & 58 & & \\
\hline
 & $\ell_R$ & $n_R$ & $\ell_L$ & $n_L$ & \\
 & 8 & 8 & 0 & 0 & \\ \hline
\end{tabular}
}
\end{center}
\label{tab2a}
\end{table}
\begin{table}[h]
\begin{center}
{\small
\begin{tabular}{cccccc} \hline
\multicolumn{6}{c}{(Ala)$_{10}$}\\ \hline
Run&\multicolumn{2}{c}{State}&
\multicolumn{3}{c}{$E$ (kcal/mol)} \\
6&\multicolumn{2}{c}{Coil}&
\multicolumn{3}{c}{$-1.5$} \\
\hline
Residue & $\phi$ & $\psi$ & $\chi$ & & \\
1 & 120 & $-58$ & 57 & & \\
2 & $-83$ & 72 & $-178$ & & \\
3 & $-65$ & 110 & $-64$ & & \\
4 & $-95$ & $-34$ & $-58$ & & \\
5 & $-153$ & 176 & 63 & & \\
6* & $-57$ & $-49$ & $-59$ & & \\
7* & $-71$ & $-50$ & $-60$ & & \\
8 & $-83$ & 74 & $-178$ & & \\
9* & $-82$ & $-22$ & 65 & & \\
10* & $-68$ & $-42$ & 177 & & \\
\hline
 & $\ell_R$ & $n_R$ & $\ell_L$ & $n_L$ & \\
 & 2, 2 & 4 & 0 & 0 & \\ \hline
\end{tabular}
}
\end{center}
\label{tab2b}
\end{table}
\newpage
Table~2. (continued) \\
\begin{table}[h]
\begin{center}
{\small
\begin{tabular}{cccccc} \hline
\multicolumn{6}{c}{(Val)$_{10}$}\\ \hline
Run&\multicolumn{2}{c}{State}&
\multicolumn{3}{c}{$E$ (kcal/mol)} \\
9&\multicolumn{2}{c}{R-Helix}&
\multicolumn{3}{c}{2.6} \\
\hline
Residue & $\phi$ & $\psi$ & $\chi^1$ & $\chi^{2,1}$ & $\chi^{2,2}$ \\
1 & 4 & $-26$ & 62 & $-62$ & $-73$ \\
2* & $-71$ & $-27$ & 77 & $-59$ & $-58$ \\
3* & $-63$ & $-29$ & 172 & 51 & $-58$ \\
4* & $-81$ & $-24$ & $-178$ & 51 & $-54$ \\
5* & $-81$ & $-35$ & 173 & $-67$ & 163 \\
6* & $-61$ & $-31$ & 162 & 51 & 54 \\
7* & $-86$ & $-50$ & 174 & 51 & $-64$ \\
8 & $-94$ & 81 & $-175$ & 56 & 60 \\
9* & $-83$ & $-43$ & 177 & 64 & 61 \\
10 & $-94$ & 122 & 177 & $-70$ & $-58$ \\
\hline
 & $\ell_R$ & $n_R$ & $\ell_L$ & $n_L$ & \\
 & 6, 1 & 7 & 0 & 0 & \\ \hline
\end{tabular}
}
\end{center}
\label{tab2d}
\end{table}
\begin{table}[h]
\begin{center}
{\small
\begin{tabular}{cccccc} \hline
\multicolumn{6}{c}{(Val)$_{10}$}\\ \hline
Run&\multicolumn{2}{c}{State}&
\multicolumn{3}{c}{$E$ (kcal/mol)} \\
10&\multicolumn{2}{c}{Coil}&
\multicolumn{3}{c}{4.0} \\
\hline
Residue & $\phi$ & $\psi$ & $\chi^1$ & $\chi^{2,1}$ & $\chi^{2,2}$ \\
1 & 42 & $-51$ & $-177$ & 57 & 65 \\
2 & $-102$ & $-52$ & $-179$ & 178 & $-54$ \\
3* & $-88$ & $-57$ & $-177$ & $-63$ & $-54$ \\
4 & $-102$ & 86 & $-178$ & 60 & 62 \\
5 & $-101$ & 8 & 77 & 71 & $-67$ \\
6* & $-85$ & $-53$ & 176 & $-66$ & 178 \\
7* & $-89$ & $-54$ & 178 & 179 & $-58$ \\
8 & $-96$ & 89 & $-172$ & 179 & $-51$ \\
9* & $-75$ & $-54$ & 177 & 52 & 173 \\
10 & $-98$ & 148 & 180 & 173 & 67 \\
\hline
 & $\ell_R$ & $n_R$ & $\ell_L$ & $n_L$ & \\
 & 2, 1, 1 & 4 & 0 & 0 & \\ \hline
\end{tabular}
}
\end{center}
\label{tab2f}
\end{table}
\newpage
Table~2. (continued) \\
\begin{table}[h]
\begin{center}
{\small
\begin{tabular}{cccccc} \hline
\multicolumn{6}{c}{(Gly)$_{10}$}\\ \hline
Run&\multicolumn{2}{c}{State}&
\multicolumn{3}{c}{$E$ (kcal/mol)} \\
6&\multicolumn{2}{c}{R-Helix}&
\multicolumn{3}{c}{0.1} \\
\hline
Residue & $\phi$ & $\psi$ & & & \\
1 & 96 & 168 & & & \\
2* & $-77$ & $-22$ & & & \\
3* & $-66$ & $-51$ & & & \\
4* & $-83$ & $-39$ & & & \\
5 & $-93$ & 60 & & & \\
6 & 164 & $-48$ & & & \\
7* & $-89$ & $-58$ & & & \\
8* & $-70$ & $-42$ & & & \\
9\# & 87 & 52 & & & \\
10 & 164 & $-124$ & & & \\
\hline
 & $\ell_R$ & $n_R$ & $\ell_L$ & $n_L$ & \\
 & 3, 2 & 5 & 1 & 1 & \\ \hline
\end{tabular}
}
\end{center}
\label{tab2h}
\end{table}
\begin{table}[h]
\begin{center}
{\small
\begin{tabular}{cccccc} \hline
\multicolumn{6}{c}{(Gly)$_{10}$}\\ \hline
Run&\multicolumn{2}{c}{State}&
\multicolumn{3}{c}{$E$ (kcal/mol)} \\
10&\multicolumn{2}{c}{L-Helix}&
\multicolumn{3}{c}{$-6.0$} \\
\hline
Residue & $\phi$ & $\psi$ & & & \\
1 & 132 & $-175$ & & & \\
2\# & 70 & 47 & & & \\
3\# & 66 & 36 & & & \\
4\# & 71 & 46 & & & \\
5\# & 65 & 46 & & & \\
6\# & 62 & 44 & & & \\
7\# & 68 & 34 & & & \\
8 & $-92$ & $-46$ & & & \\
9 & 170 & $-46$ & & & \\
10 & $-157$ & $-35$ & & & \\
\hline
 & $\ell_R$ & $n_R$ & $\ell_L$ & $n_L$ & \\
 & 0 & 0 & 6 & 6 & \\ \hline
\end{tabular}
}
\end{center}
\label{tab2hp}
\end{table}
\newpage
Table~2. (continued) \\
\begin{table}[h]
\begin{center}
{\small
\begin{tabular}{cccccc} \hline
\multicolumn{6}{c}{(Gly)$_{10}$}\\ \hline
Run&\multicolumn{2}{c}{State}&
\multicolumn{3}{c}{$E$ (kcal/mol)} \\
8&\multicolumn{2}{c}{Coil}&
\multicolumn{3}{c}{$-8.0$} \\
\hline
Residue & $\phi$ & $\psi$ & & & \\
1 & 111 & 67 & & & \\
2 & $-178$ & $-172$ & & & \\
3 & $-60$ & 136 & & & \\
4\# & 65 & 35 & & & \\
5 & 96 & $-73$ & & & \\
6 & $-153$ & 39 & & & \\
7 & $-95$ & $-71$ & & & \\
8 & $-84$ & 77 & & & \\
9 & 145 & $-116$ & & & \\
10 & $-88$ & 66 & & & \\
\hline
 & $\ell_R$ & $n_R$ & $\ell_L$ & $n_L$ & \\
 & 0 & 0 & 1 & 1 & \\ \hline
\end{tabular}
}
\end{center}
\label{tab2i}
\end{table}
\newpage
Table~3.\\
\begin{table}[h]
\begin{center}
{\small
\begin{tabular}{cccccc} \hline
\multicolumn{6}{c}{(Ala)$_{10}$}\\ \hline
\multicolumn{3}{c}{State}&
\multicolumn{3}{c}{$E$ (kcal/mol)} \\
\multicolumn{3}{c}{R-Helix}&
\multicolumn{3}{c}{$-9.7$} \\
\hline
Residue & $\phi$ & $\psi$ & $\chi$ & & \\
1  & 68 & 158 & 64 & & \\
2*  & $-67$ & $-35$ & 63 & & \\
3*  & $-68$ & $-40$ & $-60$ & & \\
4*  & $-67$ & $-38$ & 60 & & \\
5*  & $-68$ & $-38$ & $-178$ & & \\
6*  & $-68$ & $-40$ & $-59$ & & \\
7*  & $-67$ & $-36$ & $-55$ & & \\
8*  & $-71$ & $-35$ & $-177$ & & \\
9*  & $-71$ & $-40$ & $-59$ & & \\
10  & $-152$ & 106 & 56 & & \\
\hline
 & $\ell_R$ & $n_R$ & $\ell_L$ & $n_L$ & \\
 & 8 & 8 & 0 & 0 & \\ \hline
\end{tabular}
}
\end{center}
\label{tab3a}
\end{table}

\begin{table}[h]
\begin{center}
{\small
\begin{tabular}{cccccc} \hline
\multicolumn{6}{c}{(Ala)$_{10}$}\\ \hline
\multicolumn{3}{c}{State}&
\multicolumn{3}{c}{$E$ (kcal/mol)} \\
\multicolumn{3}{c}{Coil}&
\multicolumn{3}{c}{2.7} \\
\hline
Residue & $\phi$ & $\psi$ & $\chi$ & & \\
1  & 43 & 166 & 60 & & \\
2*  & $-69$ & $-27$ & $-56$ & & \\
3*  & $-75$ & $-36$ & $-63$ & & \\
4  & $-74$ & $-16$ & $-171$ & & \\
5*  & $-87$ & $-39$ & $-45$ & & \\
6*  & $-66$ & $-36$ & $-67$ & & \\
7  & $-82$ & $-11$ & 50 & & \\
8\#  & 56 & 51 & $-49$ & & \\
9  & $-110$ & $-46$ & $-62$ & & \\
10  & $-156$ & $-12$ & $-176$ & & \\
\hline
 & $\ell_R$ & $n_R$ & $\ell_L$ & $n_L$ & \\
 & 2, 2 & 4 & 1 & 1 & \\ \hline
\end{tabular}
}
\end{center}
\label{tab3b}
\end{table}

\newpage
Table~3. (continued) \\
\begin{table}[h]
\begin{center}
{\small
\begin{tabular}{cccccc} \hline
\multicolumn{6}{c}{(Val)$_{10}$}\\ \hline
\multicolumn{3}{c}{State}&
\multicolumn{3}{c}{$E$ (kcal/mol)} \\
\multicolumn{3}{c}{R-Helix}&
\multicolumn{3}{c}{1.8} \\
\hline
Residue & $\phi$ & $\psi$ & $\chi^1$ & $\chi^{2,1}$ & $\chi^{2,2}$ \\
1  & 29 & $-52$ & 178 & 51 & $-175$ \\
2*  & $-65$ & $-29$ & 72 & $-174$ & 173 \\
3*  & $-70$ & $-32$ & 166 & 54 & $-57$ \\
4*  & $-77$ & $-36$ & 176 & 178 & $-65$ \\
5*  & $-60$ & $-48$ & 163 & 172 & $-68$ \\
6*  & $-71$ & $-52$ & 174 & 55 & $-178$ \\
7  & $-102$ & 87 & $-174$ & $-64$ & $-174$ \\
8*  & $-71$ & $-36$ & 173 & $-69$ & 53 \\
9  & $-86$ & 90 & 179 & 52 & 71 \\
10  & $-85$ & 133 & 172 & $-72$ & $-65$ \\
\hline
 & $\ell_R$ & $n_R$ & $\ell_L$ & $n_L$ & \\
 & 5, 1 & 6 & 0 & 0 & \\ \hline
\end{tabular}
}
\end{center}
\label{tab3c}
\end{table}

\begin{table}[h]
\begin{center}
{\small
\begin{tabular}{cccccc} \hline
\multicolumn{6}{c}{(Val)$_{10}$}\\ \hline
\multicolumn{3}{c}{State}&
\multicolumn{3}{c}{$E$ (kcal/mol)} \\
\multicolumn{3}{c}{Coil}&
\multicolumn{3}{c}{6.6} \\
\hline
Residue & $\phi$ & $\psi$ & $\chi^1$ & $\chi^{2,1}$ & $\chi^{2,2}$ \\
1  & 68 & 164 & $-66$ & 64 & $-61$ \\
2*  & $-84$ & $-26$ & 178 & 166 & $-63$ \\
3  & $-90$ & 99 & $-176$ & $-64$ & $-52$ \\
4  & $-126$ & 30 & $-63$ & 179 & $-70$ \\
5*  & $-76$ & $-31$ & 176 & 170 & $-56$ \\
6  & $-81$ & $-16$ & $-178$ & 168 & $-58$ \\
7*  & $-82$ & $-39$ & 172 & $-71$ & 56 \\
8*  & $-80$ & $-50$ & 171 & 47 & $-62$ \\
9  & $-99$ & 90 & $-174$ & $-62$ & $-176$ \\
10  & $-87$ & 114 & $-179$ & $-61$ & 179 \\
\hline
 & $\ell_R$ & $n_R$ & $\ell_L$ & $n_L$ & \\
 & 2, 1, 1 & 4 & 0 & 0 & \\ \hline
\end{tabular}
}
\end{center}
\label{tab3d}
\end{table}

\newpage
Table~3. (continued) \\
\begin{table}[h]
\begin{center}
{\small
\begin{tabular}{cccccc} \hline
\multicolumn{6}{c}{(Gly)$_{10}$}\\ \hline
\multicolumn{3}{c}{State}&
\multicolumn{3}{c}{$E$ (kcal/mol)} \\
\multicolumn{3}{c}{R-Helix}&
\multicolumn{3}{c}{$-5.2$} \\
\hline
Residue & $\phi$ & $\psi$ & & & \\
1  & 115 & $-74$ & & & \\
2*  & $-68$ & $-36$ & & & \\
3*  & $-68$ & $-32$ & & & \\
4*  & $-71$ & $-45$ & & & \\
5*  & $-65$ & $-32$ & & & \\
6*  & $-78$ & $-56$ & & & \\
7  & 159 & 68 & & & \\
8  & $-174$ & $-106$ & & & \\
9  & $-164$ & $-86$ & & & \\
10  & $-70$ & $-81$ & & & \\
\hline
 & $\ell_R$ & $n_R$ & $\ell_L$ & $n_L$ & \\
 & 5 & 5 & 0 & 0 & \\ \hline
\end{tabular}
}
\end{center}
\label{tab3e}
\end{table}

\begin{table}[h]
\begin{center}
{\small
\begin{tabular}{cccccc} \hline
\multicolumn{6}{c}{(Gly)$_{10}$}\\ \hline
\multicolumn{3}{c}{State}&
\multicolumn{3}{c}{$E$ (kcal/mol)} \\
\multicolumn{3}{c}{L-Helix}&
\multicolumn{3}{c}{$-6.7$} \\
\hline
Residue & $\phi$ & $\psi$ & & & \\
1  & $-22$ & $-53$ & & & \\
2  & $-61$ & 95 & & & \\
3\#  & 65 & 34 & & & \\
4\#  & 76 & 35 & & & \\
5\#  & 70 & 45 & & & \\
6\#  & 75 & 39 & & & \\
7\#  & 69 & 31 & & & \\
8*  & $-87$ & $-38$ & & & \\
9  & 147 & $-33$ & & & \\
10  & $-161$ & $-50$ & & & \\
\hline
 & $\ell_R$ & $n_R$ & $\ell_L$ & $n_L$ & \\
 & 1 & 1 & 5 & 5 & \\ \hline
\end{tabular}
}
\end{center}
\label{tab3f}
\end{table}
\newpage
Table~3. (continued) \\
\begin{table}[h]
\begin{center}
{\small
\begin{tabular}{cccccc} \hline
\multicolumn{6}{c}{(Gly)$_{10}$}\\ \hline
\multicolumn{3}{c}{State}&
\multicolumn{3}{c}{$E$ (kcal/mol)} \\
\multicolumn{3}{c}{Coil}&
\multicolumn{3}{c}{$-7.6$} \\
\hline
Residue & $\phi$ & $\psi$ & & & \\
1  & $-9$ & 66 & & & \\
2  & 57 & $-106$ & & & \\
3*  & $-72$ & $-28$ & & & \\
4*  & $-71$ & $-30$ & & & \\
5  & $-85$ & $-60$ & & & \\
6  & $-179$ & 68 & & & \\
7  & 75 & $-68$ & & & \\
8  & $-130$ & 37 & & & \\
9  & $-81$ & 78 & & & \\
10  & 79 & 74 & & & \\
\hline
 & $\ell_R$ & $n_R$ & $\ell_L$ & $n_L$ & \\
 & 2 & 2 & 0 & 0 & \\ \hline
\end{tabular}
}
\end{center}
\label{tab3fp}
\end{table}
\newpage
Table~4.\\

\begin{center}
\begin{tabular}{cccccc} \hline
Peptide & $T$ & $<n>$ & $<\ell>$ & $s$ & $\sigma$ \\
\hline
(Ala)$_{10}$ & 250 & 7.8(0.4) & 7.7(0.7) & 1.6(0.1) & 0.14(0.01) \\
 & 300 & 7.5(0.6) & 7.2(1.0) & 1.5(0.2) & 0.13(0.02) \\
 & 350 & 6.8(0.8) & 6.3(1.2) & 1.3(0.2) & 0.13(0.02) \\
 & 400 & 5.5(0.9) & 4.7(1.0) & 1.1(0.2) & 0.14(0.02) \\
 & 450 & 3.5(0.7) & 2.8(0.7) & 0.76(0.12) & 0.18(0.05) \\
 & 500 & 2.1(0.5) & 1.6(0.4) & 0.49(0.10) & 0.27(0.07) \\
\hline
(Ala)$_{15}$ & 250 & 12.8(0.1) & 12.7(0.2) & 1.7(0.1) & 0.077(0.01) \\
 & 300 & 12.6(0.2) & 12.2(0.4) & 1.6(0.1) & 0.072(0.01) \\
 & 350 & 12.2(0.3) & 11.3(0.8) & 1.5(0.1) & 0.072(0.02) \\
 & 400 & 11.4(0.4) & 9.8(1.0) & 1.4(0.1) & 0.076(0.06) \\
 & 450 & 8.6(1.2) & 6.4(1.1) & 1.1(0.1) & 0.092(0.01) \\
 & 500 & 5.5(2.1) & 3.5(1.7) & 0.77(0.24) & 0.17(0.08) \\
\hline
(Ala)$_{20}$ & 250 & 17.5(0.3) & 15.9(0.9) & 1.7(0.1) & 0.051(0.01) \\
 & 300 & 17.2(0.4) & 14.9(1.4) & 1.6(0.1) & 0.051(0.01) \\
 & 350 & 16.7(0.5) & 13.2(1.5) & 1.5(0.1) & 0.052(0.01) \\
 & 400 & 15.9(0.5) & 11.1(1.5) & 1.4(0.1) & 0.053(0.01) \\
 & 450 & 14.6(0.7) & 8.8(1.2) & 1.3(0.1) & 0.058(0.01) \\
 & 500 & 9.7(1.0) & 4.3(0.2) & 0.98(0.1) & 0.091(0.02) \\
\hline
(Val)$_{10}$ & 250 & 3.9(1.3) & 2.3(0.5) & 0.73(0.25) & 0.47(0.12) \\
 & 300 & 3.4(1.2) & 2.0(0.5) & 0.60(0.22) & 0.57(0.30) \\
 & 350 & 2.8(1.0) & 1.7(0.4) & 0.45(0.17) & 0.72(0.31) \\
 & 400 & 2.3(0.9) & 1.4(0.5) & 0.36(0.18) & 0.85(0.40) \\
 & 450 & 2.0(0.8) & 1.3(0.4) & 0.30(0.17) & 0.89(0.39) \\
 & 500 & 1.7(0.8) & 1.1(0.4) & 0.26(0.16) & 0.89(0.36) \\
\hline
(Gly)$_{10}$ & 250 & 0.84(0.10) & 1.3(0.1) & 0.24(0.03) & 0.31(0.7) \\
 & 300 & 0.81(0.03) & 1.2(0.1) & 0.20(0.14) & 0.38(0.4) \\
 & 350 & 0.76(0.07) & 1.2(0.1) & 0.16(0.09) & 0.46(0.27) \\
 & 400 & 0.68(0.09) & 1.1(0.1) & 0.14(0.05) & 0.48(0.14) \\
 & 450 & 0.59(0.07) & 1.1(0.1) & 0.12(0.02) & 0.47(0.09) \\
 & 500 & 0.52(0.05) & 1.1(0.1) & 0.11(0.02) & 0.45(0.05) \\
\hline
\end{tabular}
\end{center}

\newpage
Table~5.\\

\begin{center}
\begin{tabular}{ccccc} \hline
Peptide & $T$ & $\Delta G$ & $\Delta H$ & $T\Delta S$ \\
\hline
(Ala)$_{10}$ & 250 & $-3.9(0.8)$ & $-8.1(2.1)$ & $-4.2(2.1)$ \\
 & 300 & $-3.1(0.7)$ & $-7.6(2.0)$ & $-4.5(2.0)$ \\
 & 350 & $-2.3(0.6)$ & $-9.0(1.4)$ & $-6.7(1.4)$ \\
 & 400 & $-1.2(0.6)$ & $-11.2(1.2)$ & $-10.0(1.4)$ \\
 & 450 & 0.1(0.6) & $-11.9(1.5)$ & $-12.0(1.6)$ \\
 & 500 & 1.4(0.6) & $-11.2(1.3)$ & $-12.6(1.7)$ \\
\hline
(Ala)$_{15}$ & 250 & $-11.9(1.0)$ & $-25.2(2.2)$ & $-13.3(2.4)$ \\
 & 300 & $-9.4(0.8)$ & $-23.6(2.7)$ & $-14.2(2.1)$ \\
 & 350 & $-7.0(0.7)$ & $-24.1(2.4)$ & $-17.1(2.4)$ \\
 & 400 & $-4.5(0.9)$ & $-24.3(3.2)$ & $-19.8(3.3)$ \\
 & 450 & $-2.2(1.3)$ & $-19.9(4.5)$ & $-17.7(4.7)$ \\
 & 500 & $-0.7(1.5)$ & $-13.3(6.4)$ & $-12.6(6.6)$ \\
\hline
(Ala)$_{20}$ & 250 & $-22.1(0.6)$ & $-47.9(2.3)$ & $-25.8(2.3)$ \\
 & 300 & $-17.2(0.6)$ & $-44.8(2.7)$ & $-27.6(2.7)$ \\
 & 350 & $-12.9(0.4)$ & $-42.1(2.9)$ & $-29.2(2.9)$ \\
 & 400 & $-8.8(0.3)$ & $-40.4(2.2)$ & $-31.6(2.2)$ \\
 & 450 & $-4.9(0.4)$ & $-39.4(0.8)$ & $-34.5(0.9)$ \\
 & 500 & $-1.5(0.3)$ & $-29.3(3.3)$ & $-27.8(3.3)$ \\
\hline
(Val)$_{10}$ & 250 & $-0.4(0.2)$ & $-2.1(0.7)$ & $-1.7(0.3)$ \\
 & 300 & 0.0(0.5) & $-3.0(0.7)$ & $-3.0(0.7)$ \\
 & 350 & 0.6(0.9) & $-4.1(1.5)$ & $-4.7(1.6)$ \\
 & 400 & 1.3(1.1) & $-4.4(2.1)$ & $-5.7(2.3)$ \\
 & 450 & 2.1(1.5) & $-4.4(2.5)$ & $-6.5(3.8)$ \\
 & 500 & 2.8(2.0) & $-4.6(3.4)$ & $-7.4(5.4)$ \\
\hline
(Gly)$_{10}$ & 250 & 2.0(1.0) & 0.3(1.5) & $-1.7(1.8)$ \\
 & 300 & 2.4(0.8) & $-0.3(1.1)$ & $-2.7(1.4)$ \\
 & 350 & 2.9(0.8) & $-1.1(0.9)$ & $-4.0(1.1)$ \\
 & 400 & 3.6(0.8) & $-1.8(1.6)$ & $-5.4(1.9)$ \\
 & 450 & 4.3(0.9) & $-2.4(2.0)$ & $-6.7(2.6)$ \\
 & 500 & 5.0(1.0) & $-3.0(1.8)$ & $-8.0(2.6)$ \\
\hline
\end{tabular}
\end{center}

\newpage
\noindent
{\Large Figure Captions:}\\
Fig.~1: Conformations with the lowest potential energy for each
homo-oligomer obtained from multicanonical annealing runs and
one regular multicanonical production run with 200,000 MC sweeps.
Stereoscopic views of the backbone structure are shown for (Ala)$_{10}$
  (a), (Val)$_{10}$ (b), (Gly)$_{10}$ (c), and for the lowest-energy
left-handed $\alpha$-helix obtained for (Gly)$_{10}$ (d).  In (a) and
(b) residues from 2 to 8 and from 2 to 6 are respectively helical.
The conformation in (c) is in the coil state.  Residues from 3 to 7
are helical in (d).\\
Fig.~2: Probability distribution $P (E)$ of the energy
as a function of temperature $T$ for (Ala)$_{10}$ (a), (Val)$_{10}$ (b),
and (Gly)$_{10}$ (c) obtained by one regular
multicanonical production run with 200,000 MC sweeps.\\
Fig.~3: Probability distribution $P (E)$ of the energy for T =270 K
as obtained
 by a canonical simulation ({\it can}) of 200,000 MC sweeps and a
 multicanonical simulation ({\it mul}) of 200,000 MC sweeps for
(Ala)$_{10}$ (a) and (Ala)$_{20}$ (b).\\
Fig.~4: Specific heat as a function of temperature for  (Ala)$_{10}$,
  (Val)$_{10}$, and (Gly)$_{10}$.  The values were
 calculated  from one regular multicanonical production run of
200,000 MC sweeps.\\
Fig.~5: Specific heat as a function of temperature for (Ala)$_N$
with $N=10,$ 15, and 20.
The values were calculated from one regular
  multicanonical production run of 200,000 MC sweeps.\\
{Fig.~6: Average total energy
$E_{tot} = E_{C}+E_{HB}+E_{LJ}+E_{tor}$ ($\diamondsuit$)
and averages  of its component terms, Coulomb energy $E_{C}$ ($+$),
hydrogen-bond energy $E_{HB}$ ($\Box$),
 Lennard-Jones energy $E_{LJ}$ ($X$), and torsion energy
$E_{tor}$ ($\Delta$)
 as a function of temperature $T$.  The results are separately shown for
 (Ala)$_{10}$ (a), (Val)$_{10}$ (b), (Gly)$_{10}$ (c), and
 (Ala)$_{20}$ (d).\\
Fig.~7: Average helicity $<n> \over N$ as a function of temperature $T$
 for (Ala)$_{10}$, (Val)$_{10}$, and (Gly)$_{10}$. All values were
 calculated  from one regular multicanonical production run of 200,000
MC.\\
Fig.~8: Average helicity $<n> \over N$ as a function of temperature $T$
 for (Ala)$_N$ with $N=10,$ 15, and 20.  The values were
 calculated  from one regular multicanonical production run of 200,000
MC sweeps.\\
Fig.~9: Helix propagation parameter $s$ (a) and nucleation
 parameter $\sigma$ (b) of the Zimm-Bragg model as
 a function of temperature $T$ for (Ala)$_N$ with $N=10,$ 15, and 20.
 The values were
 calculated  from one regular multicanonical production run of 200,000
MC sweeps.\\
Fig.~10: Percent helicity as a function of residue number for
 (Ala)$_{10}$, (Val)$_{10}$, and (Gly)$_{10}$ at $T=250$ K (a) and at
 $T=350$ K (b). The values were
 calculated from one regular multicanonical production run of 200,000
MC sweeps.\\

\end{document}